\documentclass[onecolumn,final]{elsart3p}
\usepackage{epsfig}

\usepackage{amssymb}
\usepackage{amsfonts}
\usepackage{dcolumn}
\journal{Nuclear Physics A}

\begin{document}

\begin{frontmatter}



\title{How the Pauli principle governs the decay of three-cluster systems}


\author{Yu. A. Lashko},
\ead{lashko@univ.kiev.ua}
\author{G. F. Filippov}
\address{Bogolyubov Institute for Theoretical Physics, 14-b Metrolohichna str., 03680, Kiev, Ukraine}

\begin{abstract}
New approach to the problem of multichannel continuum spectrum of three-cluster systems composed of an
$s$-cluster and two neutrons is suggested based on the discrete representation of a complete basis of allowed
states of the multiparticle harmonic oscillator. The structure of the eigenfunctions and behavior of the
eigenvalues of the three-cluster norm kernel are analyzed. Classification of the eigenvalues of the three-cluster
systems with the help of eigenvalues of the two-body subsystem is suggested. Asymptotic boundary conditions for a
three-cluster wave function in the continuum consistent with the requirements of the Pauli principle are
established. Such asymptotic behavior corresponds rather to subsequent decay of the three-cluster system than to
the so-called "democratic decay" associated with the hyperspherical harmonics. The $^3$H$+n+n$ configuration of
the $^5$H nucleus is considered in detail.
\end{abstract}

\end{frontmatter}

\section{Introduction}
\label{intro}

During last years the superheavy hydrogen isotope $^5$H has become an object of numerous experimental
\cite{Korsheninnikov,Golovkov,Stepantsov,Chulkov,Gurov} and theoretical
\cite{Shul'gina,Descouvemont,Arai,Fedorov,Vasilevsky} researches. The overwhelming majority of these
investigations aimed at finding the energy and width of the $^5$H resonance states. Unfortunately, even for the
$^5$H ground-state the experimental values for energy and width are significantly distinct and vary from
$E_R=1.8\pm0.1$ MeV with $\Gamma\leq0.5$ MeV \cite{Golovkov} to $E_R=5.5\pm0.2$ MeV with $\Gamma=5.4\pm0.5$ MeV
\cite{Gurov}. Detailed discussion of the available experimental data is given in Ref. \cite{Grigorenko} with the
conclusion that it is very difficult to find a non-contradictory scenario for all experimental data. Theoretical
predictions are also different: in Refs. \cite{Arai,Fedorov,Vasilevsky} a resonance $J^\pi=1/2^+$ is observed at
$E_R\simeq1.5$ MeV, while in Refs. \cite{Shul'gina,Descouvemont} nearly twice that energy was obtained. As for
the model, the most commonly used approaches treat the $^5$H nucleus as a three-cluster system composed of a
triton and two neutrons. Such assumption is justified by the fact that $^3$H is bound by 8.48 MeV \cite{Tilley},
whereas $^2$n and $^4$H subsystems are known to be unbound. Mainly, different three-cluster models fall into two
groups -- macroscopic and microscopic. In macroscopic models clusters are considered to be structureless
particles and cluster-cluster interactions are approximated by some local potentials which are fitted to
reproduce relevant data on the cluster-cluster systems. As for the Pauli exclusion principle, it is usually
simulated either with an additional repulsive potential between clusters or with orthogonalizing pseudopotential
containing the operators of projection onto the forbidden states. The first approach was used in Refs.
\cite{Shul'gina,Fedorov}, where the blocking of the Pauli-forbidden states was realized by introducing a
repulsive core in the s-wave $^3$H-neutron potentials. However, the choice of such a repulsive potential is quite
ambiguous and within this approximation a complete and accurate exclusion of the forbidden states is not ensured.
Furthermore, macroscopic three-cluster model rests on the assumption that cluster-cluster interactions are not
affected by the presence of the third cluster, but this is by no means always the case. As regards the
elimination of the Pauli-forbidden states with the help of orthogonalizing pseudopotential \cite{Kukulin}, it
could result in exclusion of some Pauli-allowed states. For example, in Ref. \cite{Fujiwara} it was shown that
some of the states of $3\alpha$-system obtained by the above-mentioned method could be regarded as spurious,
which they were not. Moreover, the latter states correspond to the most important shell-model configurations of
the $^{12}$C nucleus. Hence, the Pauli-allowed space should be carefully defined, in order not to exclude the
dominant components for the realistic description of the three-cluster systems. For this purpose, one can define
the three-cluster forbidden states referring to the harmonic oscillator wave functions of the microscopic
two-cluster subsystem, as was suggested in Ref. \cite{Fujiwara}. This idea rested on the hypothesis that the
antisymmetric three-cluster wave function should be orthogonal to the forbidden states of any two-cluster
subsystem \cite{Horiuchi}. This method seems to cope well with the task of accurate removal of the
Pauli-forbidden states in a three-cluster system. However, it does not allow to distinguish the Pauli-allowed
functions belonging to the same SU(3)-multiplet, i.e. the well-known problem of SU(3)-degeneracy of three-cluster
states could not be resolved within such approach. Moreover, the elimination of the Pauli-forbidden states does
not exhaust all exchange effects. Essential part of such effects is directly relevant to the eigenvalues of the
antisymmetrization operator. The latter eigenvalues are not identical and determine the realization probability
of the corresponding Pauli-allowed basis states in the wave function of the cluster system. Involvement of the
eigenvalues of the antisymmetrization operator in the Schr\"{o}dinger equation leads to changing the relative
kinetic energy as clusters approach each other. Consequently, clusters are shown to experience an effective
repulsion or attraction arising from the kinetic energy operator modified by the Pauli principle \cite{PRC}. Such
an effective interaction substantially affects the dynamics of the cluster-cluster interaction and can, on
occasion, produce resonance behavior of the scattering phase or even a bound state in compound nuclear system
\cite{PEPAN}. As was shown in Refs. \cite{PRC,PEPAN}, the largest eigenvalues correspond to those basis states
that are dominant in the discrete-spectrum states of the binary cluster system and in continuum states of small
above-threshold energy. Also in Ref. \cite{YaF07} it was observed that the probability of the presence of a
cluster configuration in the lowest basis function for a binary cluster system is proportional to the eigenvalue
of the isolated configuration. Hence,  the eigenvalues of the Pauli-allowed states contain a wealth of
information on a compound system composed of clusters. Meanwhile, the eigenvalues, along with corresponding
Pauli-allowed states, depend only on the assumed internal cluster functions, not on the cluster-cluster
potential, etc. Up to our knowledge, analysis of the structure of the eigenfunctions of the antisymmetrizer and
behavior of its eigenvalues has never been performed for three-cluster systems, although we believe that it could
help to establish some important laws that govern a three-cluster decay of nuclei composed of $s$-clusters.

The main difficulty in studies of resonances in a three-cluster system consists in the formulation of the correct
asymptotic boundary conditions for a wave function in the continuum. Such boundary conditions should ensure
continuous transition from the region of small intercluster distances, where the Pauli exclusion principle is of
first importance, to the asymptotic region where the scattering matrix elements are produced. Among currently
available microscopic studies of $^5$H \cite{Descouvemont,Arai,Vasilevsky}, an explicit wave function
representation of the scattering states have been employed only in Ref. \cite{Vasilevsky}. In Ref.
\cite{Descouvemont} the method of analytic continuation in the coupling constant was used for study of resonance
states in $^5$H, while in Ref. \cite{Arai} the complex scaling method was applied for the same purpose. It should
be noted that although authors of \cite{Vasilevsky} performed calculations within the same model as we do, namely
within an Algebraic Version of the resonating-group method (AVRGM) \cite{AVRGM}, our approaches essentially
differ in such aspects. To classify the three-cluster states, and enumerate the channels in the three-cluster
continuum, hyperspherical harmonics were used in Ref. \cite{Vasilevsky}, while we define the Pauli-allowed
harmonic oscillator basis functions in the Fock-Bargmann space and classify them with the use of the SU(3)
symmetry indices. The latter seems more appropriate, because the second-order Casimir operator of the SU(3) group
commutes with the operator of permutation of the nucleon position vectors. Hence, the SU(3) symmetry indices
naturally appear as the quantum numbers of the eigenfunctions of the antisymmetrization operator. The
hypermomentum, contrastingly, can not serve as quantum number of Pauli-allowed basis functions, because the Pauli
exclusion principle mixes the states with different values of hypermomentum. Hence, restriction for the maximum
value of hypermomentum $K$ included in calculation $K\leq K_{max}$ leads to spoiling the Pauli-allowed basis
functions corresponding to the number of oscillator quanta $\nu>K_{max}$, and this effect is enhanced with
increasing $\nu$. In support of this conjecture it was observed in Ref. \cite{Baye} that an accurate description
of the three-body asymptotics requires bases with large hypermomenta.

In the present paper we have shown that correct asymptotic boundary conditions can be found employing a complete
basis of the Pauli-allowed harmonic-oscillator states (classified with the use of the SU(3) symmetry indices and
defined in the Fock-Bargmann space) along with their eigenvalues. Asymptotic behavior of basis functions
consistent with the requirements of the Pauli principle gives an indication of possible decay channels of a
three-cluster nucleus and allows us to specify the most important decay channels. Such asymptotic behavior
correspond rather to subsequent decay of the three-cluster system than to the so-called "democratic decay"
associated with the hyperspherical harmonics, which are widely used for the description of three-cluster systems
(see Refs. \cite{Shul'gina,Fedorov,Vasilevsky}).

In Section \ref{sec:2} the theoretical grounds of the AVRGM are outlined and the structure of the norm kernel for
the number of three-cluster systems composed of $s$-cluster and two neutrons is discussed. Section \ref{sec:3} is
devoted to the analysis of the eigenvalues and eigenfunctions of the norm kernel of the ${^A}{\rm X}={^{A-2}}{\rm
X}+n+n$ systems ($A\leq6$). The asymptotic behavior of the coefficients of the expansion of the three-cluster
continuum wave function in the SU(3) basis is established in Section \ref{sec:4}. In Section \ref{sec:5} the
Pauli-allowed states of the $^5$H$=^3$H$+n+n$ system are considered in detail. The most important decay channels
of the $^5$H nucleus are specified and illustrated with the results of numerical calculations. Concluding remarks
are given in Section \ref{sec:6}. Definition and some properties of the Kravchuk polynomials are given in
Appendix \ref{app:1}.

\section{Theoretical background and norm kernel}
\label{sec:2}

Following the resonating-group method (RGM) \cite{RGM}, it will hereafter be supposed that the considered nuclear
systems consist of three clusters. An RGM wave function is built in the form of an antisymmetrized product of
cluster internal wave functions and a wave function of their relative motion. The internal wave functions of the
clusters are fixed\footnote{Here we shall assume the intrinsic cluster wave functions to be the simplest
functions of a translation-invariant shell model}, and the wave function of relative motion of the clusters,
which depends only on two Jacobi vectors of the considered three-cluster system, is found by solving an
integro-differential equation. The latter is obtained by substitution of the RGM wave function into the
Schr\"{o}dinger equation followed by integration with respect to single-particle coordinates. The
integro-differential equation can be transformed into the set of linear equations by expanding the wave function
of the cluster relative motion into the compete basis of the Pauli-allowed harmonic oscillator states, as the
AVRGM suggests. Another important simplification can be achieved by transformation from the coordinate space to
the space of complex generator parameters (the Fock--Bargmann space \cite{Barg}), in which basis functions are of
an especially simple form and are expressed via powers of complex vectors. Thus, the wave functions of the
considered discrete representation are reduced to power series with an infinitely large convergence radius. The
validity of this statement is indicated by the fact that all the wave functions in the Fock-Bargmann space are
entire and analytic, and, therefore, series of these functions over powers of complex vectors converge in any
finite region of the complex space.

First and foremost, the AVRGM calls for the construction of the complete basis of the Pauli-allowed harmonic
oscillator states and their classification. This is accomplished by solving the eigenvalue and eigenfunction
problem for the norm kernel, i.e., the overlap integral of the two Slater determinants composed of the
single-particle orbitals:
\begin{equation}
\label{norm_kernel_def} I(\{{\bf S}_j\},\{{\bf R}_j\})=\int \Phi(\{{\bf S}_j\},{\bf r})\Phi(\{{\bf R}_j\},{\bf
r})d\tau.
\end{equation}
Here integration is performed over all single-particles vectors, $\{{\bf R}_j\}$ (or $\{{\bf S}_j\}$) identifies
the set of three complex vectors  determining position of the center-of-mass of clusters in the Fock-Bargmann
space. For the spatial part of the single-particle wave functions we used the modified Bloch-Brink orbitals:
\begin{eqnarray*}
\phi({\bf r}_i)={1\over\pi^{3/4}} \exp\left(-{1\over2}{\bf r}^2_i+\sqrt{2}({\bf R}_j \cdot{\bf r}_i)-
{1\over2}{\bf R}_j^2\right),~~i\in A_j,
\end{eqnarray*}
where $A_j$ is the number of nucleons in the $j$th cluster. Each of these orbitals is an eigenfunction of the
coordinate operator $\hat{{\bf r}}_i$:
\begin{eqnarray*} \hat{\bf r}_i={1\over\sqrt{2}}\left({\bf
R}_j+\vec\nabla_{{\bf R}_j}\right);~~{\bf R}_j={\vec{\xi}_j+i\vec{\eta}_j\over\sqrt{2}}.
\end{eqnarray*}
which is defined in the Fock-Bargmann space and corresponds to the eigenvalue ${\bf r}_i.$ $\vec{\xi}_j$ and
$\vec{\eta}_j$ are vectors of coordinate and momentum, respectively. At the same time, orbital $\phi({\bf r}_i)$
is the kernel of the integral transform from the coordinate representation to the Fock-Bargmann representation
\cite{Barg} and the generating function for the harmonic-oscillator basis \cite{HorGen}. The center-of-mass
motion are factored out (and dropped out from now on) by transition from generator parameters $\{{\bf R}_j\}$ to
the Jacobi vectors:
\begin{eqnarray*}
{\bf R}_{cm}={1\over\sqrt{A}}\left(A_1{\bf R}_1+A_2{\bf R}_2+A_3{\bf R}_3\right),
\end{eqnarray*}
\begin{eqnarray*}
{\bf a}=\sqrt{{A_1(A_2+A_3)\over A}}\Bigg({\bf R}_1-{A_2{\bf R}_2+A_3{\bf R}_3\over {A_2+A_3}}\Bigg), ~~{\bf
b}=\sqrt{{A_2A_3\over{A_2+A_3}}}({\bf R}_2-{\bf R}_3).
\end{eqnarray*}
As the result, the overlap of two Slater determinants composed of the modified Bloch-Brink orbitals generates a
complete basis of the Pauli-allowed harmonic-oscillator functions along with their eigenvalues:
\begin{eqnarray*}
I(\{{\bf S}_j\},\{{\bf R}_j\})\Rightarrow I({\bf a},{\bf b};\tilde{\bf a},\tilde{\bf
b})=\sum_n\Lambda_n\Psi_n({\bf a},{\bf b})\Psi_n(\tilde{\bf a},\tilde{\bf b}).
\end{eqnarray*}
Functions $\Psi_n$ are defined in the Fock-Bargmann space and orthonormalized with the Bargmann measure $d\mu_B$:
\begin{eqnarray*}
d\mu_B=\exp\{-({\bf a}\tilde{{\bf a}})\}\,{d\vec{\xi_a}d\vec{\eta_a}\over(2\pi)^3}\,\exp\{-({\bf b}\tilde{{\bf
b}})\}\,{d\vec{\xi_b}d\vec{\eta_b}\over(2\pi)^3},
\end{eqnarray*}
$n$ stands for the set of quantum numbers of basis functions. Quantum numbers and the structure of functions
$\Psi_n$ will be discussed in the next section.

Examine next the explicit expressions for the norm kernels of those three-cluster systems which comprised of an
$s$-cluster and two neutrons, such as $^6$He=$^4$He$+n+n$, $^5$H=$^3$H$+n+n$, $^4n=^2n+n+n$, $^3n=n+n+n$,
subsequently referred to in this article as ${^A}{\rm X}={^{A-2}}{\rm X}+n+n,~A\leq6.$ The norm kernels for all
these systems can be written in general form:
\begin{eqnarray}
I_{S_{(nn)}=0}&=&\exp({\bf a}\tilde{\bf a})\cosh({\bf b}\tilde{\bf b})-\exp({\bf
a}_+\tilde{\bf a})\cosh({\bf b}_+\tilde{\bf b})-\nonumber\\
& &-\exp({\bf a}_-\tilde{\bf a})\cosh({\bf b}_-\tilde{\bf b})+\exp\{({\bf a}_0\tilde{\bf a})\},
\label{norm_singlet}
\end{eqnarray}
\begin{eqnarray}
I_{S_{(nn)}=1}&=&\exp({\bf a}\tilde{\bf a})\sinh({\bf b}\tilde{\bf b})-\exp({\bf a}_+\tilde{\bf a})\sinh({\bf
b}_+\tilde{\bf b})-\nonumber\\
& &-\exp({\bf a}_-\tilde{\bf a})\sinh({\bf b}_-\tilde{\bf b})+\exp\{({\bf a}_0\tilde{\bf a})\}.
\label{norm_triplet}
\end{eqnarray}
Here $S_{(nn)}$ designates the spin of the two-neutron subsystem. At that the so-called "T-tree" of the Jacobi
coordinates is considered, i.e., vector ${\bf b}$ describes relative distance between two neutrons, while vector
${\bf a}$ indicates the distance from the remaining cluster to the center of mass of the two-neutron subsystem.
Note that $I_{S_{(nn)}=0}$ is symmetric with respect to the permutation of two valence neutrons, while
$I_{S_{(nn)}=1}$ is antisymmetric as it must.

Each term in Eqs. (\ref{norm_singlet}) and (\ref{norm_triplet}) corresponds to a certain permutation of identical
nucleons. The first exponent is associated with the identity permutation, the second and the third exponents
appeared as the result of permutation of a valence neutron and a neutron belonging to the $s$-cluster, while the
last term relates to the simultaneous permutation of two pairs of identical neutrons:
\begin{eqnarray}
\label{permut} {\bf a}_\pm=\left(1-{t\over2}\right){\bf a}\pm{1\over2}\sqrt{t}\,{\bf b},~~{\bf
b}_\pm=\pm{1\over2}\sqrt{t}\,{\bf a}+{1\over2}{\bf b},~~{\bf a}_0=\left(1-t\right){\bf a}.
\end{eqnarray}
Finally, the parameter $t$ is positive and equal to the square of tangent of the angle of rotation from one
Jacobi tree to another. $t$ takes the value $3/2$ for $^6$He nucleus, $5/3$ for the $^5$H, $t=2$ for tetraneutron
and $t=3$ for three-neutron system. As can readily be observed, parameter $t$ decreases with increasing the
number of nucleons in $s$-cluster.

The eigenvalues of the norm kernel do not depend on the choice of the Jacobi tree, whereas the structure of the
eigenfunctions does. Since the Pauli-allowed basis states of the ${^A}{\rm X}={^{A-2}}{\rm X}+n+n$ systems take
the simplest form in the "T-tree", it precisely this tree is  best suited to the construction and analysis of
these states.

\section{Eigenvalues and eigenfunctions of the norm kernel of the ${^A}{\rm X}={^{A-2}}{\rm X}+n+n$ systems}
\label{sec:3}

Let us discuss now the set of quantum numbers of the Pauli-allowed states $\Psi_n({\bf a},{\bf b}).$ For the
three-cluster systems considered here, $n$ includes the number of oscillator quanta $\nu$, the indices
$(\lambda,\mu)$ of their SU(3) symmetry, the additional quantum number $k$ if there are several differing
$(\lambda,\mu)$ multiplets, the orbital momentum $L$ and its projection $M$ and, if necessary, one more
additional quantum number $\alpha_{L}$. The latter is needed to label the states with the same $L$ in a given
$(\lambda,\mu)$ multiplet. The spin of the two-neutron subsystem $S_{(nn)}$ also is an integral of motion  for as
long as the spin-orbital interaction is switched off. It is well known that the diagonalization of the norm
kernel requires the basis to be labeled with the quantum indices $(\lambda,\mu)$ of irreducible representations
of the SU(3) group \cite{HorFud}. The eigenvalues of the norm kernel depend on the total number of the oscillator
quanta and $(\lambda,\mu)$, and do not depend on the angular momenta of the basis states. We shall restrict our
consideration to the states with zero orbital momentum $L=0$ and positive parity. Hence, the number of oscillator
quanta should be even and equal $2\nu$, quantum numbers $L$ and $M$ will be dropped from now on.

By definition, the states $\Psi_n({\bf a},{\bf b})$ are the eigenfunctions of the antisymmetrization operator
$\hat{A}$:
\begin{eqnarray*}
\hat{A}\Psi_{(2\nu-4\mu,2\mu)_{k}}^{S_{(nn)}}=\Lambda_{(2\nu-4\mu,2\mu)_{k}}^{S_{(nn)}}\Psi_{(2\nu-4\mu,2\mu)_{k}}^{S_{(nn)}}.
\end{eqnarray*}
In two-cluster systems the eigenvalues of antisymmetrizer limit to unity at $\nu\rightarrow\infty$ and the
deviations from the unity are due to the Pauli exclusion principle. Contrastingly, eigenvalues of three-cluster
systems tend to eigenvalues of a two-cluster subsystem with increasing the number of oscillator quanta $\nu$:
\begin{eqnarray*}
\lim_{\nu-2\mu\to\infty}\Lambda^{{^A}{\rm X}={^{A-2}}{\rm
X}+n+n}_{(2\nu-4\mu,2\mu)_{k}}\rightarrow\lambda_{k}^{{^{A-1}}{\rm X}={^{A-2}}{\rm X}+n}.
\end{eqnarray*}
Coincidence of the eigenvalues of the three-cluster norm kernels with those of the two-cluster subsystem norm
kernels in the limit of the large number of quanta has been already pointed out in \cite{Kato}, where $3\alpha$
and $^{16}$O$+2\alpha$ systems have been considered. However, no relation with the additional quantum number $k$
labeling the states with the same SU(3)-indices has been established. This subject will be pursued further in the
next section. Now let us proceed with the Pauli-allowed basis functions
$\Psi_{(2\nu-4\mu,2\mu)_{k}}^{S_{(nn)}}({\bf a},{\bf b}).$

In the Fock-Bargmann space the latter functions are the superpositions of the eigenfunctions of the second-order
Casimir operator:
\begin{eqnarray*}
\Psi_{(2\nu-4\mu,2\mu)_{k}}^{S_{(nn)}=0}({\bf a},{\bf
b})=\sum_{m=\mu}^{\nu-\mu}D^{2m-2\mu}_{(2\nu-4\mu,2\mu)_{k}}\psi_{(2\nu-4\mu,2\mu)}^{2m-2\mu}({\bf a},{\bf b}),
\end{eqnarray*}
\begin{eqnarray*}
\Psi_{(2\nu-4\mu,2\mu)_{k}}^{S_{(nn)}=1}({\bf a},{\bf
b})=\sum_{m=\mu}^{\nu-\mu-1}D^{2m-2\mu+1}_{(2\nu-4\mu,2\mu)_{k}}\psi_{(2\nu-4\mu,2\mu)}^{2m-2\mu+1}({\bf a},{\bf
b}),
\end{eqnarray*}
Functions $\psi_{(2\nu-4\mu,2\mu)}^{2m-2\mu}({\bf a},{\bf b})$ and $\psi_{(2\nu-4\mu,2\mu)}^{2m-2\mu+1}({\bf
a},{\bf b})$ correspond to the same total number of oscillator quanta $2\nu$, but differ in the number of quanta
along vectors ${\bf a}$ and ${\bf b}$. That is the reason why the three-cluster states are always degenerate,
even though only $s$-clusters involved, and the degree of the SU(3) degeneracy increases drastically with
increasing the number of quanta. It was found in Ref. \cite{jmp} that the eigenfunctions of the Casimir operator
with powers of vectors ${\bf a}$ and ${\bf b}$ being fixed are expressible in terms of hypergeometric functions
$_2F_1(\alpha,\beta;\gamma; Z)$, with the variable
$$Z={[{\bf ab}]^2\over {\bf a}^2{\bf b}^2}.$$
Namely,
\begin{eqnarray*}
\psi^{2m-2\mu}_{(2\nu-4\mu,2\mu)}({\bf a},{\bf b})&=&N^{2m-2\mu}_{(2\nu-4\mu,2\mu)} [{\bf ab}]^{2\mu}{\bf
a}^{2\nu-2m-2\mu}{\bf
b}^{2m-2\mu}\times\\
& &\times{_2}F_1\left(-\nu+m+\mu,-m+\mu;-\nu+2\mu+{1\over2};Z\right),
\end{eqnarray*}
\begin{eqnarray*}
\psi^{2m-2\mu+1}_{(2\nu-4\mu,2\mu)}({\bf a},{\bf b})&=&N^{2m-2\mu+1}_{(2\nu-4\mu,2\mu)} [{\bf ab}]^{2\mu}({\bf
ab}){\bf
a}^{2\nu-2m-2\mu-2}{\bf b}^{2m-2\mu}\times\\
& &\times{_2}F_1\left(-\nu+m+\mu+1,-m+\mu;-\nu+2\mu+{1\over2};Z\right).
\end{eqnarray*}
Here $N^{2m-2\mu}_{(2\nu-4\mu,2\mu)}$ and $N^{2m-2\mu+1}_{(2\nu-4\mu,2\mu)}$ are the normalization coefficients
of the corresponding basis functions.

The norm kernel $I_{S_{(nn)}}$ can be written down as a sum of partial norm kernels with definite values of the
number of quanta $2\nu$:
\begin{eqnarray*}
I_{S_{(nn)}=0}=\sum_{\nu=0}^{\infty}I^{2\nu}_{S_{(nn)}=0},~~
I_{S_{(nn)}=1}=\sum_{\nu=1}^{\infty}I^{2\nu}_{S_{(nn)}=1}.
\end{eqnarray*}
In the Fock--Bargmann representation, at a given $\nu$ the norm kernel $I^{2\nu}_{S_{(nn)}}$ is always
representable in the form of a sum of $SU(3)$-projected norm kernels $I^{(\lambda,\mu)}_{S_{(nn)}}$. We shall
deal with the relevant part of the norm kernel, $I^{(2\nu-4\mu,2\mu)}_{S_{(nn)}}$ and write it as
\begin{eqnarray*}
I^{(2\nu-4\mu,2\mu)}_{S_{(nn)}=0}&=&\sum_{m=\mu}^{\nu-\mu}\sum_{\tilde{m}=0}^{\nu-m-\mu}
\langle2\nu,2m-2\mu|2\nu,2m-2\mu+2\tilde{m}\rangle\times\\
& &\times \left(\psi^{2m-2\mu}_{(2\nu-4\mu,2\mu)}\tilde{\psi}^{2m-2\mu+2\tilde{m}}_{(2\nu-4\mu,2\mu)}+
\psi^{2m-2\mu+2\tilde{m}}_{(2\nu-4\mu,2\mu)}\tilde{\psi}^{2m-2\mu}_{(2\nu-4\mu,2\mu)}\right),
\end{eqnarray*}
\begin{eqnarray*}
I^{(2\nu-4\mu,2\mu)}_{S_{(nn)}=1}&=&\sum_{m=\mu}^{\nu-\mu-1}\sum_{\tilde{m}=0}^{\nu-m-\mu-1}
\langle2\nu,2m+1-2\mu|2\nu,2m+1-2\mu+2\tilde{m}\rangle\times\\
& &\times \left(\psi^{2m+1-2\mu}_{(2\nu-4\mu,2\mu)}\tilde{\psi}^{2m+1-2\mu+2\tilde{m}}_{(2\nu-4\mu,2\mu)}+
\psi^{2m+1-2\mu+2\tilde{m}}_{(2\nu-4\mu,2\mu)}\tilde{\psi}^{2m+1-2\mu}_{(2\nu-4\mu,2\mu)}\right),
\end{eqnarray*}
where
\begin{eqnarray*}
\langle2\nu,2m-2\mu|2\nu,2m-2\mu+2\tilde{m}\rangle= \delta_{\tilde{m},0}\left(1+\delta_{m,0}\delta_{\mu,0}(1-t)^{2\nu}\right)-\\
-{2\,t^{\tilde{m}}\over4^{\nu-\mu}}\left({t-1\over
t-2}\right)^{2\mu}(t-2)^{2(\nu-m-\tilde{m})}\sqrt{(2\nu-2m-2\mu)!(2m+2\tilde{m}-2\mu)!\over(2\nu-2m-2\tilde{m}-2\mu)!(2m-2\mu)!}\times\\
\times  {_2}F_1\left(-2\nu+2m+2\mu+2\tilde{m},-2m+2\mu;2\tilde{m}+1;{t\over2-t}\right)
\end{eqnarray*}
\begin{eqnarray*}
\langle2\nu,2m+1-2\mu|2\nu,2m+1-2\mu+2\tilde{m}\rangle= \delta_{\tilde{m},0}+\\
+{2\,t^{\tilde{m}}\over4^{\nu-\mu}}\left({t-1\over
t-2}\right)^{2\mu}(t-2)^{2(\nu-m-\tilde{m})-1}\sqrt{(2\nu-2m-2\mu-1)!(2m+2\tilde{m}+1-2\mu)!\over(2\nu-2m-2\tilde{m}-1-2\mu)!(2m+1-2\mu)!}\times\\
\times  {_2}F_1\left(-2\nu+2m+2\mu+1+2\tilde{m},-2m-1+2\mu;2\tilde{m}+1;{t\over2-t}\right).
\end{eqnarray*}
Obviously, the Pauli-allowed basis functions $\Psi_{(2\nu-4\mu,2\mu)_{k}}^{S_{(nn)}}$ can be found by the
diagonalization of the norm kernels $I^{(2\nu-4\mu,2\mu)}_{S_{(nn)}}$ at a given $\nu.$ There are $\nu-2\mu+1$
different states possessing SU(3) symmetry $(2\nu-4\mu,2\mu)$, but some of them correspond to zero eigenvalues
and thus they are forbidden by the Pauli principle. The number of the latter functions depends on the system
considered. Hence, solving the eigenvalue and eigenfunction problem for the norm kernel $I({\bf a},{\bf
b};\tilde{\bf a},\tilde{\bf b})$ in the Fock-Bargmann space, we arrive at all the Pauli-allowed and the
Pauli-forbidden functions of the ${^A}{\rm X}={^{A-2}}{\rm X}+n+n$ systems given in terms of orthogonal
polynomials of a discrete variable.

As the eigenvalues $\Lambda^{{^A}{\rm X}={^{A-2}}{\rm X}+n+n}_{(2\nu-4\mu,2\mu)_{k}}$ of the norm kernel approach
limit values $\lambda_{k}^{{^{A-1}}{\rm X}={^{A-2}}{\rm X}+n}$, corresponding eigenvectors
$\Psi_{(2\nu-4\mu,2\mu)_{k}}$ take simple analytical form:
\begin{eqnarray}
\label{func_as_S0} \Psi_{(2\nu-4\mu,2\mu)_{k}}^{S_{(nn)}=0}({\bf a},{\bf
b})\rightarrow{1\over\sqrt{2}}\left(\psi_{(2\nu-4\mu,2\mu)}^{k-2\mu}({\bf a}_1,{\bf
b}_1)+\psi_{(2\nu-4\mu,2\mu)}^{k-2\mu}({\bf a}_2,{\bf b}_2)\right),
\end{eqnarray}
\begin{eqnarray}
\label{func_as_S1} \Psi_{(2\nu-4\mu,2\mu)_{k}}^{S_{(nn)}=1}({\bf a},{\bf
b})\rightarrow{1\over\sqrt{2}}\left(\psi_{(2\nu-4\mu,2\mu)}^{k-2\mu}({\bf a}_1,{\bf
b}_1)-\psi_{(2\nu-4\mu,2\mu)}^{k-2\mu}({\bf a}_2,{\bf b}_2)\right).
\end{eqnarray}
Such asymptotical behavior takes place with the proviso that $\nu\gg k,$ but as will be exemplified with $^5$H
system, the fulfilment of condition $\nu\geq k+5$ would be ample. Note that $k$ is the natural integer that falls
in the range $2\mu\leq k\leq \nu.$

Here ${\bf a}_1,{\bf b}_1$ and ${\bf a}_2,{\bf b}_2$ are the Jacobi vectors of the so-called "Y-tree", with
vector ${\bf b}_1$ $({\bf b}_2)$ describing relative distance between one of the valence neutron and
${^{A-2}}{\rm X}$-cluster. Correspondingly, vector ${\bf a}_1$ $({\bf a}_2)$ determines position of the remaining
neutron relative to the center-of-mass of the ${^{A-1}}{\rm X}$ subsystem. Jacobi vectors of the "T-tree" are
related to those of the "Y-tree" via unitary transformation:
\begin{eqnarray*}
{\bf a}=\cos\alpha\,{\bf a}_{1,2}+\sin\alpha\,{\bf b}_{1,2};{\bf b}=\mp\sin\alpha\,{\bf
a}_{1,2}\pm\cos\alpha\,{\bf b}_{1,2}.
\end{eqnarray*}
Naturally, function $\Psi_{(2\nu-4\mu,2\mu)_{k}}^{S_{(nn)}=0}$ is symmetric with respect to the interchange of
the vectors $({\bf a}_1,{\bf b}_1)$ and $({\bf a}_2,{\bf b}_2)$, while $\Psi_{(2\nu-4\mu,2\mu)_{k}}^{S_{(nn)}=1}$
is antisymmetric, in so far as such operation corresponds to the permutation of the valence neutrons.


The remarkable feature of asymptotical relations (\ref{func_as_S0}) and (\ref{func_as_S1}) lies in the fact that
in the limit $\nu\gg k$ the expansion coefficients
$D^{2m-2\mu}_{(2\nu-4\mu,2\mu)_{k}},D^{2m-2\mu+1}_{(2\nu-4\mu,2\mu)_{k}}$ can be identified with the Kravchuk
polynomials of a discrete variable \cite{Kravchuk}:
\begin{eqnarray*}
D^{2m-2\mu}_{(2n-4\mu,2\mu)_{k}}\rightarrow\sqrt{2}\,{\cal K}^{(p)}_{k-2\mu}(2m-2\mu){\sqrt{\rho_{2m-2\mu}}\over
d_{k-2\mu}},
\end{eqnarray*}
\begin{eqnarray*}
D^{2m-2\mu+1}_{(2n-4\mu,2\mu)_{k}}\rightarrow\sqrt{2}\,{\cal
K}^{(p)}_{k-2\mu}(2m-2\mu+1){\sqrt{\rho_{2m-2\mu+1}}\over d_{k-2\mu}},~~p=\sin^2\alpha.
\end{eqnarray*}
The Kravchuk polynomials ${\cal K}^{(p)}_{k-2\mu}(m-2\mu)$ are specified on the interval $2\mu\leq
m\leq2\nu-2\mu$ and orthogonal with weighting function $\rho_{m-2\mu}$ and norm $d_{k-2\mu}$. An explicit
expression for the Kravchuk polynomials, as well as some of their basic properties, are given in Appendix
\ref{app:1}. Here it is worth noting that the Kravchuk polynomials are the discrete analog of the Hermitian
polynomials.

So, the degree of the Kravchuk polynomials serves as the additional quantum number of the SU(3) degenerate
three-cluster states. The Pauli-allowed basis states $\Psi_{(2\nu-4\mu,2\mu)_{k}}^{S_{(nn)}}$ can be arranged
into branches and families, with all the states of a particular branch having common symmetry index $\mu$ and
overlapping generously with corresponding asymptotic function (\ref{func_as_S0}) or (\ref{func_as_S1}), but
differing in value of the first index $\lambda$. The eigenvalues belonging to a given branch tend to the same
limit value $\lambda_{k}^{{^{A-1}}{\rm X}}$ with the number of quanta increasing. The branches which share limit
eigenvalues are combined in the family of the eigenstates, which thus is completely determined by the degree $k$
of the corresponding Kravchuk polynomial.

Evidence for the importance of a particular family of the Pauli-allowed states  can be found by analyzing the
behavior of their eigenvalues with the increasing the number of oscillator quanta $\nu.$ In the Introduction
mention was made of the fact that the change in the kinetic energy of relative motion under the effect of the
Pauli exclusion principle leads to an effective interaction of the nuclei as they approach each other. As is
shown in Refs. \cite{PRC,PEPAN}, the character of such interaction in two-cluster systems is determined by the
behavior of the eigenvalues of the antisymmetrization operator. They are nonnegative, since they are proportional
to the probability of the realization of the corresponding allowed basis state, and tend to unity with increasing
distance between the nuclei involved. In the case where the eigenvalues of the antisymmetrization operator tend
to unity from below, the effective interaction proves to be repulsive, but, if these eigenvalues tend to unity
from above, one can say that there arises attraction induced by the Pauli exclusion principle. As indicated
earlier, the eigenvalues $\Lambda^{{^A}{\rm X}={^{A-2}}{\rm X}+n+n}_{(2\nu-4\mu,2\mu)_{k}}$ of the three-cluster
norm kernel approach the eigenvalues of the two-cluster subsystem $\lambda_{k}^{{^{A-1}}{\rm X}}$ with increasing
the number of quanta $\nu.$ It would appear reasonable that families of the states with
$\Lambda_{(2\nu-4\mu,2\mu)_{k}}<\lambda_{k}$ would be suppressed at small $\nu,$ which can naturally be
considered as a demonstration of the action of effective forces of repulsion for small intercluster distances.
Contrastingly, those families which characterized by $\Lambda_{(2\nu-4\mu,2\mu)_{k}}>\lambda_{k}$ should become
more favorable at small $\nu$, which can be considered as effective attraction. The validity of these assumptions
for the $^5$H nucleus will be demonstrated in Section \ref{sec:5}.

It is worth noting that the effective interaction caused by the change in the kinetic energy of the relative
motion of clusters under the effect of the antisymmetrization operator arises only between objects whose internal
energy may change when they approach each other\footnote{Here, we imply the interaction of nuclei consisting of
nucleons.}. Therefore, no such interaction would appear between two point-like nucleons. By this reason, the
Pauli-allowed states in $^3$n=n+n+n system will most likely have the asymptotic behavior different from
(\ref{func_as_S0}),(\ref{func_as_S1}).

\section{Asymptotic equations for the expansion coefficients}
\label{sec:4}

We seek the wave function of the considered three-cluster system in the form of expansion over the SU(3) basis of
the Pauli-allowed states
\begin{eqnarray}
\label{wf_exp} \Upsilon_{\kappa\,(E)}({\bf a},{\bf b})=\sum_n \sqrt{\Lambda_n}C_n^{\kappa\,(E)}\Psi_n({\bf
a},{\bf b}).
\end{eqnarray}
Expansion coefficients both of the discrete eigenstates with energy $E_\kappa=-\kappa^2/2<0$ and of the continuum
eigenstates $\left\{C_{n}(E)\right\}$ with energy $E>0$ are found by solving a set of linear equations
\begin{eqnarray}
\label{eq} \sum_{n'}\langle n|\hat{H}|n'\rangle C_{n'}-E \Lambda_nC_{n}=0.
\end{eqnarray}
In Ref. \cite{Fewbody2} we have shown that in order to set the asymptotic boundary conditions for the expansion
coefficients of a two-cluster wave function in the SU(3) basis, a basis with a different set of quantum numbers
(the angular-momentum coupled basis) is required. The states of the latter basis are labeled with angular momenta
of the clusters and of their relative motion. The equations (\ref{eq}) for the expansion coefficients in the
SU(3) basis remain coupled even in the asymptotic region, whereas the set of corresponding equations in the
angular-momentum coupled basis is uncoupled at a large number of excitation quanta. The transformation between
the two is defined through a unitary matrix and can be found with the use of the integration technique in the
Fock--Bargmann space developed by the present authors. However, the norm kernel of binary cluster system has a
diagonal form only in the representation of the basis of its eigenfunctions, i.e., in the representation of the
SU(3)-basis. A unitary transform of the basis disrupts the diagonal form of the norm kernel due to the difference
between the eigenvalues for different SU(3) representations. This would not be the case if all the eigenvalues
were equal. The nature of this breaking is that, unlike the functions of the SU(3) basis, those of the
angular-momentum coupled basis are not eigenfunctions of the antisymmetrization operator and, therefore, are not
invariant with respect to a permutation of the nucleons. The permutation mixes the angular-momentum coupled basis
functions with the same number of quanta. However, as the latter increases, the degree of mixing decreases
exponentially, and at large values of the number of oscillator quanta the norm kernel becomes practically
diagonal in the angular-momentum coupled basis as well. Hence, in this region the asymptotic behavior of the
expansion coefficients in the angular-momentum coupled basis can be defined  (they are expressed in terms of
Hankel functions of the first and second kind, and the scattering $S$-matrix elements) and related to that of the
expansion coefficients in the SU(3) basis.

Contrastingly to the eigenvalues of the two-cluster norm kernels, those of the three-cluster norm kernel remain
distinct even in the limit $\nu\to\infty$. Existence of different limit eigenvalues $\lambda_k$ reflects the
possibility for two of three clusters to be close to each other and far apart from the third cluster. Owing to
this the unitary transformation from the SU(3)-basis to the other one is rendered possible only within a
particular family of the Pauli-allowed states. Any transformation, which involves the states belonging to
different families (for example, transformation to the three-cluster hyperspherical harmonics), would disrupt the
diagonal form of the norm kernel and, hence, is inappropriate to the occasion.

Turning back to the asymptotic SU(3) basis functions (\ref{func_as_S0}),(\ref{func_as_S1}), it might be well to
point out that the latter functions have a simple physical meaning. They reproduce relative motion of a
two-cluster subsystem $^{A-1}{\rm X}$ occurring in a ground or an excited harmonic-oscillator state and a
remaining neutron. Such asymptotic behavior gives an indication of possible decay channels of a three-cluster
nucleus and allows us to specify the most important decay channels of the nucleus under consideration.

In the limit $\nu\gg k$ instead of asymptotical SU(3) basis functions
$${1\over\sqrt{2}}\left(\psi_{(2\nu-4\mu,2\mu)}^{k-2\mu}({\bf a}_1,{\bf
b}_1)\pm\psi_{(2\nu-4\mu,2\mu)}^{k-2\mu}({\bf a}_2,{\bf b}_2)\right)$$ it is advantageous to use the angular
momentum basis functions $${1\over\sqrt{2}}\left(\phi_{2\nu,k}^l({\bf a}_1,{\bf b}_1)\pm\phi_{2\nu,k}^l({\bf
a}_2,{\bf b}_2)\right),$$ defined as follows:
\begin{eqnarray*}
\phi_{2\nu,k}^l({\bf a},{\bf b})=N_{\nu,k}^l
{\bf a}^{2\nu-k-l}{\bf b}^{k-l}({\bf
ab})^{l}\cdot{1\over2^l}\sum_{\tilde{l}=0}^{[l/2]}{(-1)^{\tilde{l}}(2l-2\tilde{l})!\over
\tilde{l}!(l-\tilde{l})!(l-2\tilde{l})!}{{\bf a}^{2\tilde{l}}{\bf b}^{2\tilde{l}}\over({\bf ab})^{2\tilde{l}}}.
\end{eqnarray*}
The states of this basis (referred to as "$l$-basis" in what follows) are labeled by the number of quanta $2\nu$,
and the angular momenta $l$ of the $^{A-1}{\rm X}$ subsystem coinciding with the angular momentum of the relative
motion of this subsystem and a remaining neutron\footnote{Recall that only the states with total angular momentum
$L=0$ are considered in this paper.}. $N_{\nu,k}^l$ is the normalization coefficient.

For $\nu\gg k$ variables in (\ref{eq}) are separated and the latter set of equations is representable in the form
of a sum of two sets of equations. One of them describes the relative motion of $^{A-1}{\rm X}$ subsystem and a
neutron:
\begin{eqnarray*}
-{1\over4}\sqrt{(2\nu-k-l)(2\nu-k+l+1)}C^{l}_{\nu-1,k}+
{1\over2}\left(2\nu-k+{3\over2}-2\tilde{E}\right)C^{l}_{\nu,k}-
\end{eqnarray*}
\begin{eqnarray}
\label{rel_dif} -{1\over4}\sqrt{(2\nu-k-l+2)(2\nu-k+l+3)}C^{l}_{\nu+1,k}=0,
\end{eqnarray}
while the other one characterizes the $^{A-1}{\rm X}=^{A-2}{\rm X}+n$ subsystem itself:
\begin{eqnarray}
-\sqrt{{\lambda_{k-2}\over\lambda_{k}}}\cdot{1\over4}\sqrt{(k-l)(k+l+1)}C^{l}_{\nu-1,k-2}+
{1\over2}\left(k+{3\over2}-2\varepsilon\right)C^{l}_{\nu,k}-\nonumber\\
\label{subsyst_dif}
-\sqrt{{\lambda_{k}\over\lambda_{k+2}}}\cdot{1\over4}\sqrt{(k-l+2)(k+l+3)}C^{l}_{\nu+1,k+2}=0.
\end{eqnarray}
We here put the nucleon mass, the Planck's constant and oscillator length equal to 1 for the sake of brevity.

Thus asymptotically the three-cluster Schr\"{o}dinger equation (\ref{eq}) can be reduced to a two-body-like
multichannel problem. Total energy $E$ is distributed between energy $\varepsilon$ of the two-cluster subsystem
$^{A-1}{\rm X}$ and energy $\tilde{E}$ of the relative motion of the above-mentioned subsystem and a remaining
neutron so that $\tilde{E}+\varepsilon=E.$ Note that Eq. (\ref{rel_dif}) corresponds to the free motion of a
neutron and $^{A-1}{\rm X}$ subsystem, while the matrix of Eq. (\ref{rel_dif}) contains the limit eigenvalues
$\lambda_k$ of the allowed states and, for this reason, is not identical to the matrix corresponding to the
kinetic energy operator of free motion of the $^{A-2}{\rm X}$ cluster and a neutron. In other words, it contains
some effective cluster-cluster interaction derived from the kinetic-energy operator modified by the Pauli
principle. It is precisely this interaction that causes the $^{A}{\rm X}$ system to decay via an intermediate
stage, i.e.,$^{A}{\rm X}\rightarrow^{A-1}{\rm X}+n\rightarrow^{A-2}{\rm X}+n+n.$ Alternative decay channel
$^{A}{\rm X}\rightarrow^{A-2}{\rm X}+^2n\rightarrow^{A-2}{\rm X}+n+n$ is much less favorable, because in this
case Eq. (\ref{subsyst_dif}) would describe just a free motion of two neutrons and, thus, a sufficiently high
energy is needed to make this decay channel open. The possibility for such decay channel to exist in $^5$H is
discussed in Section \ref{sec:5}.

From the above discussion it appears that the greater is the number of the families of the Pauli-allowed states
invoked, the better is the description of the $^{A-1}{\rm X}$ subsystem and the less is its energy $\varepsilon.$
Failure to take into account a sufficient number of families results in too localized $^{A-1}{\rm X}$ subsystem
and, hence, in too high energy $\varepsilon.$ Noteworthy also is the fact that in the limit of the large number
of oscillator quanta $\nu$ the kinetic energy operator does not couple the states belonging to the families
corresponding to even values of quantum number $k$ with those characterized by odd-numbered $k$. Such phenomenon
takes place because even and odd values of $k$ correspond to the states of the $^{A-1}{\rm X}$ subsystem with
different parity, which becomes an integral of motion at large distance between the $^{A-1}{\rm X}$ subsystem and
a neutron.

Of course, the cluster-cluster interaction generated by the nucleon-nucleon potential should also participate in
Eq. (\ref{rel_dif}) providing for realistic description of the $^{A-1}{\rm X}$ subsystem. However, inclusion of
the nucleon-nucleon interaction between nucleons belonging to different clusters could not change the structure
of the Pauli-allowed states and in a qualitative sense affect asymptotic behavior of the expansion coefficients
of the wave function over these states. Study of the effects related to the potential energy (with its exchange
part) in the microscopic analytical approach described here is under its way, and the results will be published
in a separate paper.

Difference equations (\ref{rel_dif}) and (\ref{subsyst_dif}) become Bessel equations in the limit of large $\nu$
and $k$:
\begin{eqnarray}
\label{rel_bes} \left({d^2\over dy_\nu^2}+{1\over y_\nu}{d\over
dy_\nu}+2\tilde{E}-{(2l+1)^2\over{4y_\nu^2}}\right)C^l_k(y_\nu)=0;~~y_\nu=\sqrt{4\nu-2k+3}.
\end{eqnarray}
\begin{eqnarray}
\label{subsyst_bes} \left({d^2\over dy_k^2}+{1\over y_k}{d\over
dy_k}+2\varepsilon-{(2l+1)^2\over{4y_k^2}}\right)C^l_\nu(y_k)=0;~~y_k=\sqrt{2k+3}.
\end{eqnarray}
Therefore, the asymptotic form of the expansion coefficients in the $l$-basis can be conveniently written in
terms of the Hankel functions $H^{\pm}_{l+1/2}$ and the scattering $S$-matrix elements. If the incoming wave is
in the channel characterized by angular momentum $l$ and energy $E-\varepsilon_i$ of the relative motion of
$^{A-1}{\rm X}$ subsystem and a neutron, the expansion coefficients in this channel satisfy the asymptotic
relation
\begin{eqnarray*}
C^{l}_{\nu,k}(E-\varepsilon_i)=H^{(-)}_{l+1/2}\left(\sqrt{2(E-\varepsilon_i)}\sqrt{4\nu-2k+3}\right)
H^{(-)}_{l+1/2}\left(\sqrt{2\varepsilon_i}\sqrt{2k+3}\right)-\\
-\sum_{j=1}^{k_{\rm max}} {\bf S}_{ij}H^{(+)}_{l+1/2}\left(\sqrt{2(E-\varepsilon_j)}\sqrt{4\nu-2k+3}\right)
H^{(+)}_{l+1/2}\left(\sqrt{2\varepsilon_j}\sqrt{2k+3}\right).
\end{eqnarray*}
Here index $i$ enumerates different decay channels and its maximum possible value is equal to the number of
different families considered. We emphasize that energy levels $\varepsilon_i$ of the $^{A-1}{\rm X}$ subsystem
do not necessarily belong to the discrete states, but can be continuum states as well. Even in the latter case it
is possible to indicate the most favorable decay channels. For example, the $^{A-1}{\rm X}$ subsystem may not
have any bound states, but could have a low-energy resonance. In such a case the $^{A-2}{\rm X}+n+n$
three-cluster system would most likely decay via resonance state of the $^{A-1}{\rm X}$ subsystem.

\section{The Pauli-allowed states of the $^3$H$+n+n$ system}
\label{sec:5}

Let us use a three-cluster configuration $^3$H$+n+n$ of the $^5$H nucleus with $J^\pi=1/2^+$ to illustrate the
validity of our conclusions. We shall focuss our attention on the states with positive parity, zero total orbital
momentum and zero spin of the two-neutron subsystem. For the sake of brevity the above-mentioned quantum numbers
will be dropped in what follows.

In this case the norm kernel (\ref{norm_singlet}) takes the form:
\begin{eqnarray}
I&=&\cosh({\bf a}\tilde{\bf a})\cosh({\bf b}\tilde{\bf b})-\cosh({\bf
a}_+\tilde{\bf a})\cosh({\bf b}_+\tilde{\bf b})-\nonumber\\
& &-\cosh({\bf a}_-\tilde{\bf a})\cosh({\bf b}_-\tilde{\bf b})+\cosh\{({\bf a}_0\tilde{\bf a})\},
\label{norm_singlet_5H}
\end{eqnarray}
where ${\bf a}_\pm,{\bf b}_\pm$ and ${\bf a}_0$ are defined by Eq. (\ref{permut}) with $t=5/3.$

The norm kernel (\ref{norm_singlet_5H}) have to be projected to the states with definite indices of the SU(3)
symmetry $(2\nu-4\mu,2\mu)$ and expanded over a complete set of the eigenstates $\Psi_{(2\nu-4\mu,2\mu)_{k}}({\bf
a},{\bf b})$ of the antisymmetrization operator:
\begin{eqnarray*}
I_{(2\nu-4\mu,2\mu)}&=&\sum_{k=2\mu}^{\nu}\Lambda_{(2\nu-4\mu,2\mu)_{k}}\Psi_{(2\nu-4\mu,2\mu)_{k}}.
\end{eqnarray*}
Eigenvalues of the norm kernel can be arranged into the branches and families quite as discussed in Section
\ref{sec:3}. Eigenvalues belonging to the first five families are given in Table \ref{table:1}.

\begin{table*}[htb]
\caption{\label{table:1} Eigenvalues $\Lambda_{(2\nu-4\mu,2\mu)_{k}}$ of the norm
 kernel for the system $^3$H$+n+n$}
\begin{tabular}{|cccccccccccc|}
\hline

 & $k=1$ & \multicolumn{2}{c}{$k=2$} & \multicolumn{2}{c}{$k=3$} & \multicolumn{3}{c}{$k=4$} & \multicolumn{3}{c}{$k=5$} \\

$\nu$ & $\mu=0$ & $\mu=0$ & $\mu=1$ & $\mu=0$ & $\mu=1$ & $\mu=0$ & $\mu=1$ & $\mu=2$ & $\mu=0$ & $\mu=1$ &
$\mu=2$ \\

\hline 1 & 1.8889 &  &  &  &  &  &  &  &  &  & \\

2 & 1.4780 & 0.9541  & 0.7778&  &  &  &  &  &  &  & \\

3 & 1.3487 & 0.8814 & 0.8733 & 1.0895 & 1.0649 &  &  &  &  &  &  \\

4 & 1.3344 & 0.8862 & 0.8879 & 1.0636 & 1.0434 & 1.0161 & 0.9837 & 0.9753 & & & \\

5 & 1.3334 & 0.8885 & 0.8888 & 1.0431 & 1.0377 & 0.9841 & 0.9847 & 0.9859 & 1.0139 & 1.0034 & 1.0072 \\

6 & 1.3333 & 0.8888 & 0.8888 & 1.0378 & 1.0371 & 0.9861 & 0.9870 & 0.9875 & 1.0088 & 1.0055 & 1.0048 \\

7 & 1.3333 & 0.8889 & 0.8889 & 1.0371 & 1.0370 & 0.9873 & 0.9876 & 0.9876 & 1.0056 & 1.0045 & 1.0042 \\

8 & 1.3333 & 0.8889 & 0.8889 & 1.0370 & 1.0370 & 0.9876 & 0.9876 & 0.9876 & 1.0044 & 1.0042 & 1.0041 \\

9 & 1.3333 & 0.8889 & 0.8889 & 1.0370 & 1.0370 & 0.9876 & 0.9876 & 0.9876 & 1.0042 & 1.0041 & 1.0041 \\

10 & 1.3333 & 0.8889 & 0.8889 & 1.0370 & 1.0370 & 0.9876 & 0.9876 & 0.9876 & 1.0041 & 1.0041 & 1.0041 \\

\hline
\end{tabular}
\end{table*}

One can readily see from Table \ref{table:1} that the $^5$H norm kernel eigenvalues belonging to the $k$th family
tend to the eigenvalues of a two-cluster subsystem $^4$H with increasing the number of oscillator quanta $\nu$:
\begin{eqnarray*}
\lim_{\nu-2\mu\to\infty}\Lambda^{{^5}{\rm H}={^3}{\rm
H}+n+n}_{(2\nu-4\mu,2\mu)_{k}}\rightarrow\lambda_{k}^{{^4}{\rm H}={^3}{\rm H}+n}=1-\left(-{1\over3}\right)^k.
\end{eqnarray*}
Obviously, index $k$ makes sense of the number of oscillator quanta accounted for by the $^4$H subsystem. Of
special note is the fact that $\lambda_{k=2\tilde{k}+1}>1,$ while $\lambda_{k=2\tilde{k}}<1.$ This in itself is
indicative of attraction between $^3$H and a neutron in the states with odd number of quanta $k=2\tilde{k}+1$ and
repulsion in the states with even number of quanta $k=2\tilde{k}.$ Such a conclusion can be easily verified by
analyzing the asymptotic Eq. (\ref{subsyst_dif}). Moreover, the eigenvalues
$\Lambda_{(2\nu-4\mu,2\mu)_{2\tilde{k}+1}}$ approach limit eigenvalues $\lambda_{2\tilde{k}+1}$ from above,
contrastingly to $\Lambda_{(2\nu-4\mu,2\mu)_{2\tilde{k}}}$ approaching $\lambda_{2\tilde{k}}$ from below. As was
concluded in Section \ref{sec:3}, there are strong grounds for believing that families of the states
characterized by odd values of quantum number $k$ dominates in the wave function of the $^5$H system.
\begin{table*}[htb]
\caption{\label{table:2} Overlap integrals $\int\Psi_{(2\nu-4\mu,2\mu)_{k}}\Psi_{(2\nu-4\mu,2\mu)_{k}}^{\rm
as}d\mu_B$ versus the number of quanta $\nu$ for the first five families of the Pauli-allowed states of the
$^5$H.}
\begin{tabular}{|cccccccccccc|}
\hline

 & $k=1$ & \multicolumn{2}{c}{$k=2$} & \multicolumn{2}{c}{$k=3$} & \multicolumn{3}{c}{$k=4$} & \multicolumn{3}{c}{$k=5$} \\

$\nu$ & $\mu=0$ & $\mu=0$ & $\mu=1$ & $\mu=0$ & $\mu=1$ & $\mu=0$ & $\mu=1$ & $\mu=2$ & $\mu=0$ & $\mu=1$ &$\mu=2$ \\

\hline
1     & 0.9701  &         &         &         &          &        &         &         &         &         & \\

2     & 0.9460  & 0.9924  & 1       &         &          &        &         &         &         &         & \\

3     & 0.9893  & 0.9979  & 0.9934  & 0.8076  & 0.9915   &        &         &         &         &         &  \\

4     & 0.9996  & 0.9998  & 0.9992  & 0.9007  & 0.9886   & 0.9214 & 0.9629  & 1       &         &         & \\

5     & 1.0000  & 1.0000  & 1.0000  & 0.9736  & 0.9983   & 0.9706 & 0.9900  & 0.9934  & 0.5887  & 0.9052  &  0.9915\\

6     & 1.0000  & 1.0000  & 1.0000  & 0.9981  & 0.9999   & 0.9962 & 0.9979  & 0.9992  & 0.7299  & 0.9907  &  0.9886\\

7     & 1.0000  & 1.0000  & 1.0000  & 0.9999  & 1.0000   & 0.9995 & 0.9998  & 1.0000  & 0.9294  & 0.9974  &  0.9983\\

8     & 1.0000  & 1.0000  & 1.0000  & 1.0000  & 1.0000   & 1.0000 & 1.0000  & 1.0000  & 0.9932  & 0.9996  &  0.9999\\

9     & 1.0000  & 1.0000  & 1.0000  & 1.0000  & 1.0000   & 1.0000 & 1.0000  & 1.0000  & 0.9996  & 1.0000  &  1.0000 \\

10    & 1.0000  & 1.0000  & 1.0000  & 1.0000  & 1.0000   & 1.0000 & 1.0000  & 1.0000  & 1.0000  & 1.0000   & 1.0000 \\

\hline
\end{tabular}
\end{table*}
Table \ref{table:2} illustrates the fact that the Pauli-allowed states $\Psi_{(2\nu-4\mu,2\mu)_{k}}$ take
asymptotic form
$$\Psi_{(2\nu-4\mu,2\mu)_{k}}^{\rm as}={1\over\sqrt{2}}\left(\psi_{(2\nu-4\mu,2\mu)}^{k-2\mu}({\bf a}_1,{\bf
b}_1)+\psi_{(2\nu-4\mu,2\mu)}^{k-2\mu}({\bf a}_2,{\bf b}_2)\right)$$ as $\Lambda_{(2\nu-4\mu,2\mu)_{k}}$ approach
their limit eigenvalues. In Table \ref{table:2} overlap integrals of the exact Pauli-allowed states with their
asymptotic expressions $\Psi_{(2\nu-4\mu,2\mu)_{k}}^{\rm as}$ depending on the number of quanta $2\nu$ are given
for the first five families of the $^5$H norm kernel eigenfunctions. As long as $\nu\leq5$, the states listed in
Table \ref{table:2} exhaust all possible basis functions allowed by the Pauli principle. As $\nu$ increases, new
families of states characterized by $k>5$ also appear. However, eigenstates and eigenvalues belonging to such
families are governed by the same laws as those tabulated in Table \ref{table:2}. Furthermore, states of "odd"
families, i.e., $\Psi_{(2\nu-4\mu,2\mu)_{2\tilde{k}+1}}$ appear to dominate over the states
$\Psi_{(2\nu-4\mu,2\mu)_{2\tilde{k}}}$, with small $\tilde{k}$  prevailing. The first family
$\Psi_{(2\nu,0)_{k=1}}$ plays a leading part in the wave function of the $^5$H system, followed by the families
$\Psi_{(2\nu-4\mu,2\mu)_{k=3}}$ and $\Psi_{(2\nu-4\mu,2\mu)_{k=5}}.$ Such hierarchy among different families of
the Pauli-allowed states of the $^5$H is directly related to the behavior of the corresponding eigenvalues
$\Lambda_{(2\nu-4\mu,2\mu)_{2\tilde{k}+1}}$ discussed above. An exact treatment of the antisymmetrization effects
related to the kinetic energy exclusively was shown to result in an effective attraction of the clusters in those
branches whose eigenvalues exceed their limiting values.

From Table \ref{table:1} and Table \ref{table:2} we can observe that a complete space of the Pauli-allowed basis
functions can be divided into "internal" region, where $\nu-k<5$, and asymptotic region ($\nu-k\geq5$). In the
latter region  functions $\Psi_{(2\nu-4\mu,2\mu)_{k}}$ take simple analytical form
$\Psi_{(2\nu-4\mu,2\mu)_{k}}^{\rm as}$ and correspond to the decay of the $^5$H nucleus into the $^4$H subsystem
and a neutron, with the $^4$H cluster occurring in the state with $k$ oscillator quanta. In the internal region
the Pauli-allowed states have more complicated form, which is indicative of exchange effects involving all three
clusters.

The states of the $k$th family have one more important feature, which makes small $k$ predominating. The less is
$k$, the more smooth basis functions $\Psi_{(2\nu-4\mu,2\mu)_{k}}$ are. Let us demonstrate this with the
functions belonging to the first family consisting of a single branch $\Psi_{(2\nu-4\mu,2\mu)_{k=1}}.$ The latter
branch has a remarkably simple asymptotic form:
\begin{eqnarray*}
\Psi_{(2\nu,0)_1}^{\rm as}={1\over\sqrt{2}}\cdot\sqrt{{2\nu\over(2\nu+1)!}}\left\{{\bf a}_1^{2\nu-2}({\bf
a}_1{\bf b}_1)+{\bf a}_2^{2\nu-2}({\bf a}_2{\bf b}_2)\right\},
\end{eqnarray*}
and the exact function $\Psi_{(2\nu-4\mu,2\mu)_{k=1}}$ becomes indistinguishable from the asymptotic one at
$\nu=5.$ It is easily comprehended that function $\Psi_{(2\nu,0)_1}^{\rm as}$ corresponds to the decay of the
$^5$H nucleus into the $^4$H subsystem being in the lowest oscillator shell model state and a neutron. Indeed,
$\Psi_{(2\nu,0)_1}^{\rm as}$ contains only the first power of vector ${\bf b}_1$ or ${\bf b}_2$ evidencing that
$^3$H cluster and a neutron remain at the minimal distance from each other, which is consistent with the
requirements of the Pauli exclusion principle. In turn, eigenvalues of the first family $\Lambda_{(2\nu,0)_1}$
tend to the lowest eigenvalue of the $^4$H norm kernel $\lambda_{k=1}=4/3.$ It is noteworthy also that the
orbital angular momentum $l$ of the relative motion of the neutron cluster and $^4$H subsystem, as well as the
orbital momentum of the $^4$H itself is $l=1$ for the states of the first branch.

\begin{figure}[tbh]
\includegraphics{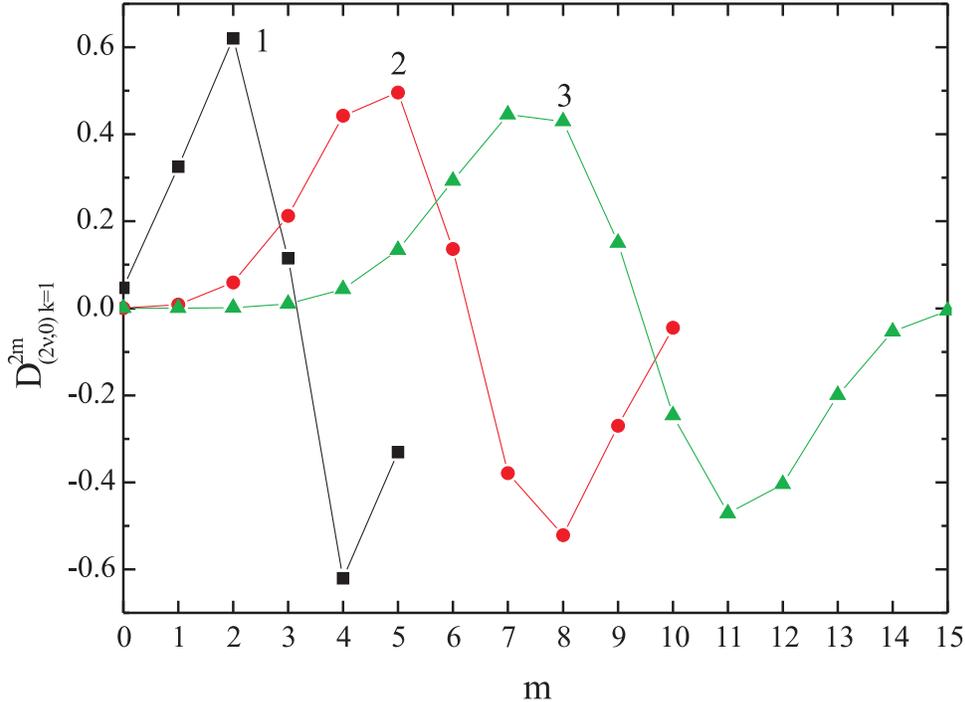}
\caption{\label{fig:1} Expansion coefficients $D^{2m}_{(2\nu,0)_{k=1}}$ of the first branch versus $m$ at
different values of the number of quanta $\nu$. Curves: \textit{1} -- $\nu=5$; \textit{2} -- $\nu=10$; \textit{3}
-- $\nu=15$.}
\end{figure}

As noted in Section \ref{sec:3}, at $\nu\geq k+5$ the expansion coefficients $D^{2m-2\mu}_{(2\nu-4\mu,2\mu)_{k}}$
of the Pauli-allowed basis functions $\Psi_{(2\nu-4\mu,2\mu)_{k}}$ can be identified with the Kravchuk
polynomials of a discrete variable $2m-2\mu$:
\begin{eqnarray*}
D^{2m-2\mu}_{(2\nu-4\mu,2\mu)_{k}}\rightarrow\sqrt{2}\,{\cal
K}^{(5/8)}_{k-2\mu}(2m-2\mu){\sqrt{\rho_{2m-2\mu}}\over d_{k-2\mu}}.
\end{eqnarray*}
Dependence of the expansion coefficients of the first branch $\Psi_{(2\nu-4\mu,2\mu)_{k=1}}$ on the number of
quanta $\nu$ is shown in Fig.~\ref{fig:1}.
At a given $\nu$ there are exists $\nu+1$ coefficients $D^{2m}_{(2\nu,0)_{k=1}}$. As expected, the latter
coefficients have one node at all $\nu$, because their asymptotic behavior is determined by the Kravchuk
polynomials of the first order. Note that with increase in $\nu$ coefficients $D^{2m}_{(2\nu,0)_{k=1}}$ become
Hermitian polynomials of the first order as follows from (\ref{Hermit}). Evidently, the less is the number of
nodes, the lower is the kinetic energy and potential energy corresponding to a given branch. Therefore, families
of the allowed states associated with the low-order Kravchuk polynomials are more favorable.

It should be pointed out that amongst eigenfunctions of the antisymmetrization operator of the $^5$H system
belonging to SU(3) representation $(2\nu,0)$ there is a nodeless branch of eigenfunctions
$\Psi_{(2\nu,0)_{k=0}}.$ However, it falls into the category of the Pauli-forbidden states and corresponds to
zero eigenvalue. Functions of this branch can be written in analytical form for a given $\nu$:
\begin{eqnarray*}
\Psi_{{(2\nu,0)}_{k=0}}=N_{{(2\nu,0)}_{k=0}}\cdot{1\over\sqrt{(2\nu+1)!}}\left\{{\bf a}_1^{2\nu}+{\bf
a}_2^{2\nu}\right\},
\end{eqnarray*}
where $N_{{(2\nu,0)}_{k=0}}$ is the normalization coefficient, which tends to $1/\sqrt{2}$ with $\nu$ increasing.
The expansion coefficients $D^{2m}_{(2\nu,0)_{k=0}}$ are given by the following expression:
\begin{eqnarray*}
D^{2m}_{(2\nu,0)_{k=0}}=\left({3\over8}\right)^\nu\sqrt{{(2\nu)!\over(2\nu-2m)!(2m)!}} \left({5\over3}\right)^m.
\end{eqnarray*}
The structure of $\Psi_{(2\nu,0)_{k=0}}$ indicates that one of the valence neutrons occupies an $s$-state in the
$^4$H subsystem, which is forbidden by the Pauli principle.

Let us discuss now possibility for another decay channel of the $^5$H to exist. At first glance it would seem
that $^5$H$\rightarrow^2n+^3$H$\rightarrow n+n+^3$H decay should occur along with the $^5$H$\rightarrow
n+^4$H$\rightarrow n+n+^3$H decay. However, the structure of the Pauli-allowed functions is not compatible with
such assumption. Basic functions corresponding to the "dineutron" decay of the $^5$H, if any, should look like
\begin{eqnarray}
\label{func_as_2n} \Psi_{(2\nu-4\mu,2\mu)_{2m}}({\bf a},{\bf b})\rightarrow\psi_{(2\nu-4\mu,2\mu)}^{2m-2\mu}({\bf
a},{\bf b}),~~\mu\leq m\leq\nu-\mu.
\end{eqnarray}
with eigenvalues
\begin{eqnarray*}
\lim_{\nu-2\mu\to\infty}\Lambda_{(2\nu-4\mu,2\mu)_{2m}}\rightarrow\lambda_{2m}^{^2n=n+n}=1.
\end{eqnarray*}
Asymptotic behavior of the simplest function relevant to the decay of the $^5$H into the dineutron being in the
lowest shell-model state and triton is of the form
\begin{eqnarray}
\label{2n0_as} \Psi_{(2\nu,0)_{m=0}}^{\rm as}={1\over\sqrt{(2\nu+1)!}}{\bf a}^{2\nu}.
\end{eqnarray}
From this it readily follows that
\begin{eqnarray*}
\lim_{\nu\to\infty}D^{2m}_{(2\nu,0)_{m=0}}\to\delta_{m,0}.
\end{eqnarray*}

In Table \ref{table:3} eigenvalues of those basis states which most closely resemble functions (\ref{2n0_as}) are
listed along with the weight of the state $\psi_{(2\nu,0)}^{m=0}({\bf a},{\bf b})$.

\begin{table}[htb]
\caption[]{\label{table:3} Eigenvalues of the candidate for the "dineutron decay" function
$\Psi_{(2\nu,0)_{m=0}}$ of the $^5$H and overlap \\$D^{m=0}_{(2\nu,0)}=\int\Psi_{(2\nu,0)_{m=0}}^{\rm
as}\Psi_{(2\nu,0)_{m=0}}d\mu_B$ versus number of quanta $\nu$.}
\begin{tabular}{|ccccccccccc|}
\hline
$\nu$ & 1 & 2 & 3 & 4 & 5 & 6 & 7 & 8 & 9 & 10\\
$\Lambda_{(2\nu,0)_{m=0}}$ & 1.8889 & 1.4781 & 1.0895 & 1.0636 & 1.0139 & 1.0087 & 1.0027 & 1.0015 & 1.0008 & 1.0003\\
$D^{m=0}_{(2\nu,0)}$ & 0.8575 & 0.7304 & 0.8407 & 0.7327 & 0.8564 & 0.8510 & 0.8141 & 0.9518 & 0.8237 & 0.9406\\
\hline
\end{tabular}
\end{table}

\begin{figure}[tbh]
\includegraphics{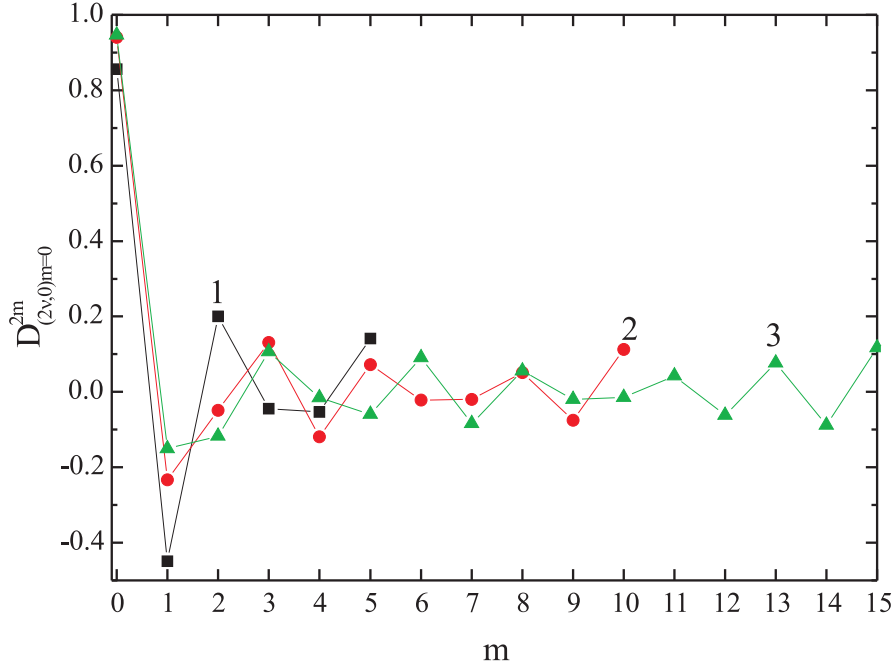}
\caption{\label{fig:2} Expansion coefficients $D^{2m}_{(2\nu,0)_{m=0}}$ of the "dineutron decay" branch versus
$m$ at different values of the number of quanta $\nu$. Curves: \textit{1} -- $\nu=5$; \textit{2} -- $\nu=10$;
\textit{3} -- $\nu=15$.}
\end{figure}

As may be inferred from this table, the dependence of the coefficient $D^{m=0}_{(2\nu,0)}$ from the number of
quanta $\nu$ is not monotone and even at $\nu=10$ the weight of the "dineutron decay" function is only $88\%$.
Note that the weight of the simplest $^4$H$+n$ function in the states of the first family runs up to $100\%$ at
$\nu=5.$ As regards the branches $\Psi_{(2\nu,0)_{2m>0}}^{\rm as}$, which describe the decay of the $^5$H into
the dineutron being in the state with $2m$ quanta and a triton, none of the Pauli-allowed states of the $^5$H has
such asymptotic behavior. Figure \ref{fig:2} depicts the expansion coefficients $D^{2m}_{(2\nu,0)_{m=0}}$ of the
branch, which is a candidate for the "dineutron decay" mode, at different values of the number of quanta $\nu.$
As evident from Fig.~\ref{fig:2}, coefficients $D^{2m}_{(2\nu,0)_{m=0}}$ have noticeable tail even at $\nu=15,$
indicating that function $\Psi_{(2\nu,0)_{m=0}}$ still does not coincide with its asymptotic expression
$\Psi_{(2\nu,0)_{m=0}}^{\rm as}.$  Furthermore, it has too many nodes and thus quite high energy is needed to
excite such a mode. All the foregoing counts in favour of the conclusion that
$^5$H$\rightarrow^2n+^3$H$\rightarrow n+n+^3$H decay is not realized. As for the states
$\Psi_{(2\nu-4\mu,2\mu)_{k}}$, which are dominated by the component $\psi_{(2\nu,0)}^{m=\mu}({\bf a},{\bf b})$,
they may be thought of as exhibiting some $nn$ correlations in the internal region $\nu-k<5.$

In regard to the "democratic decay" of the $^5$H, it is highly improbable due to the fact that all limit
eigenvalues of the $^5$H nucleus, namely $\lambda_{k}^{{^4}{\rm H}}$, are different. Distinction of limit
eigenvalues implies that there is no region where the Pauli-allowed states of the three-cluster system $^5$H
coincide with hyperspherical harmonics, which reproduce the "democratic decay".
\begin{figure}[tbh]
\includegraphics{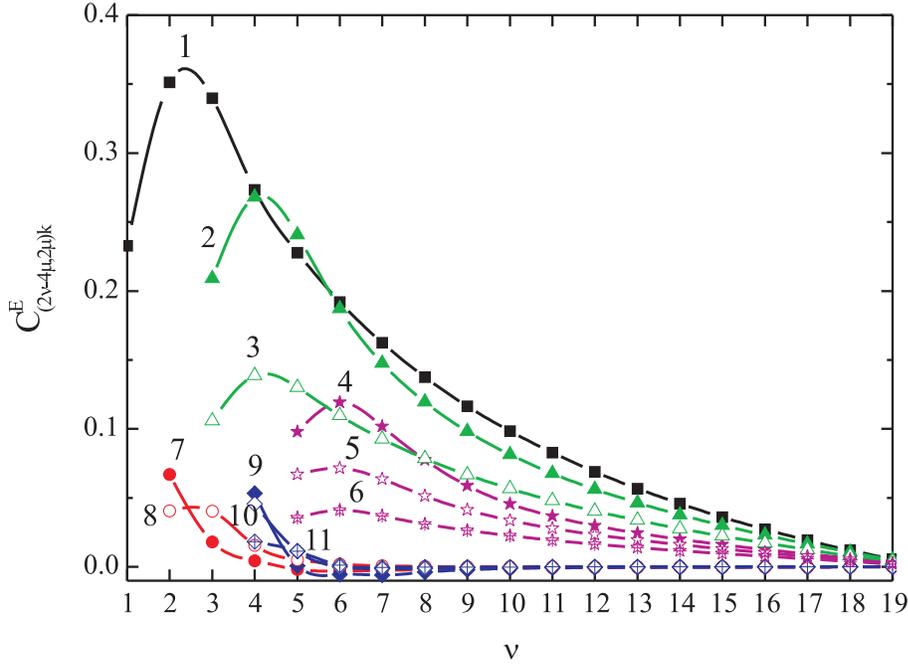}
\caption{\label{fig:3} Coefficients $C_{(2\nu-4\mu,2\mu)_{k}}^E$ of the expansion of the $^5$H wave function in
the SU(3)-basis at $E=10.67$ MeV. Curves: \textit{1} -- $(2\nu,0)_{k=1}$; \textit{2} -- $(2\nu,0)_{k=3}$;
\textit{3} -- $(2\nu-4,2)_{k=3}$; \textit{4} -- $(2\nu,0)_{k=5}$; \textit{5} -- $(2\nu-4,2)_{k=5}$; \textit{6} --
$(2\nu-8,4)_{k=5}$; \textit{7} -- $(2\nu,0)_{k=2}$; \textit{8} -- $(2\nu-4,2)_{k=2}$; \textit{9} --
$(2\nu,0)_{k=4}$; \textit{10} -- $(2\nu-4,2)_{k=4}$; \textit{11} -- $(2\nu-8,4)_{k=4}$.}
\end{figure}
Finally, Figure \ref{fig:3} demonstrates the coefficients $C_{(2\nu-4\mu,2\mu)_{k}}^E$ appearing in the expansion
of the $^5$H wave function $\Upsilon_{\kappa\,(E)}({\bf a},{\bf b})$ with energy $E=10.67$ MeV above the
$^5$H$\rightarrow^3$H$+n+n$ decay threshold in the basis of the Pauli-allowed states
$\Psi_{(2\nu-4\mu,2\mu)_{k}}.$ These coefficients have been obtained by diagonalization of the Hamiltonian, in
which only the operator of the kinetic energy of the relative motion (with its exchange part) is retained. First
five families were taken into complete account, families with $k>5$ were disregarded. Oscillator length $r_0$ was
chosen so to reproduce root-mean-square radius of the $^3$H cluster. In fact, the eigenfunctions of the kinetic
energy operator modified by the Pauli principle demonstrate the manner in which the antisymmetrization influences
the expansion coefficients $C_{(2\nu-4\mu,2\mu)_{k}}^E$. Figure \ref{fig:3} sustains our conclusion that families
characterized by odd values of $k$ contribute much more significantly to the wave function of the $^5$H than
families with even $k$ do. An hierarchy among the coefficients of different branches of a given family is
established by the behavior of their eigenvalues, namely, by the magnitude of attraction or repulsion in the
corresponding SU(3) branches. By this reason, the coefficients of the branches $(2\nu,0)_{k=1}$ take the lead,
followed by the $(2\nu,0)_{k=3}$ and $(2\nu-4,2)_{k=3}.$ Neglecting families of the states with $k=2$ and $k=4$
does not change considerably the wave function of the $^5$H and the energy $E.$ The latter takes such a large
value mainly due to the fact that energy $\varepsilon$ of the $^4$H subsystem is high enough at given
localization of the $^4$H. Of course, taking into account nucleon-nucleon interaction between $^3$H cluster and a
neutron would reduce this energy. Involvement of the "odd" families with higher $k$ also leads to better
description of the $^4$H subsystem and, hence, to lowering the energy $E.$  It should be also pointed out that
the behavior of the expansion coefficients $C_{(2\nu-4\mu,2\mu)_{k}}^E$ provides support for the predominance of
the decay $^5$H$\rightarrow n+^4$H$\rightarrow n+n+^3$H.

\section{Conclusion}
\label{sec:6}

New approach to the problem of multichannel continuum spectrum of ${^A}{\rm X}={^{A-2}}{\rm X}+n+n$ systems
($A\leq6$) is suggested based on the discrete representation of a complete basis of allowed states of the
multiparticle harmonic oscillator, which was systematized with the help of the indices of SU(3) symmetry and
defined in the Fock-Bargmann space. Proposed approach allows correct description of three-cluster systems both in
the region of small intercluster distances, where the Pauli exclusion principle is of first importance, and in
the asymptotic region where the scattering matrix elements are produced.

Careful analysis of the structure of the eigenfunctions and behavior of the eigenvalues of the three-cluster norm
kernel has been performed for the first time. A set of linear algebraic equations which generates the
Pauli-allowed states for different three-cluster systems composed of an $s$-cluster and two neutrons is written
in a general form. The latter depends on the single parameter which can be easily found for any three-cluster
system.

In the Fock-Bargmann space the Pauli-allowed states of a three-cluster system are shown to be superpositions of
the hypergeometric functions, with the expansion coefficients being orthogonal polynomials of a discrete
variable. Eigenvalues of the three-cluster system are shown to tend to eigenvalues of a two-cluster subsystem
with increasing the number of oscillator quanta $\nu$. At the same time, corresponding eigenvectors take simple
analytical form, while the expansion coefficients become the Kravchuk polynomials as the number of oscillator
quanta increases. We suggest a way of resolving the problem of the SU(3) degeneracy of the Pauli-allowed states.
A degree of the Kravchuk polynomial serves as an additional quantum number to label the states belonging to the
same SU(3) irreducible representations.

The Pauli-allowed states of a three-cluster system ${^{A-2}}{\rm X}+n+n$ can be arranged into branches and
families, with all the states of a particular branch having common SU(3)-symmetry index $\mu$, but differing in
value of the first SU(3) index $\lambda$. The eigenvalues belonging to a given branch tend to the same eigenvalue
of the two-body subsystem $\lambda_{k}^{{^{A-1}}{\rm X}}$ with the number of quanta increasing. The branches
which share limit eigenvalues are combined in the family of the eigenstates, which thus is completely determined
by the degree $k$ of the corresponding Kravchuk polynomial. Each family of the Pauli-allowed states
asymptotically corresponds to a certain binary decay channel of a three-cluster system into a two-cluster
subsystem $^{A-1}{\rm X}$ occurring in a ground or an excited harmonic-oscillator state and a remaining neutron.
Such asymptotic behavior gives an indication of possible decay channels of a three-cluster nucleus and allows us
to specify the most important decay channels of the nucleus under consideration.

The families which characterized by eigenvalues approaching limit values $\lambda_{k}$ from above are shown to be
more favorable at small $\nu$, which can be considered as effective attraction of clusters due to the exchange
effects. It is precisely this interaction that causes the $^{A}{\rm X}$ system to decay via an intermediate
stage, i.e.,$^{A}{\rm X}\rightarrow^{A-1}{\rm X}+n\rightarrow^{A-2}{\rm X}+n+n.$ Alternative decay channel
$^{A}{\rm X}\rightarrow^{A-2}{\rm X}+^2n\rightarrow^{A-2}{\rm X}+n+n$ is not realized.

It is observed that distinction of limit eigenvalues of the three-cluster norm kernel results in the absence of
such region where the Pauli-allowed states of the $^{A-2}{\rm X}+n+n$ system coincide with hyperspherical
harmonics, which reproduce the "democratic decay". Hence, instead of hyperspherical functions the
angular-momentum basis functions, which are labeled by the number of quanta, the angular momentum of the
$^{A-1}{\rm X}$ subsystem and momentum of the relative motion of this subsystem and a remaining neutron, should
be used in the limit $\nu\gg k$. As the result, asymptotically the three-cluster Schr\"{o}dinger equation can be
reduced to a two-body-like multichannel problem, while the asymptotic form of the expansion coefficients in the
$l$-basis is expressed in terms of the Hankel functions and the scattering $S$-matrix elements.

The validity of these conclusions was illustrated with the $^3$H$+n+n$ configuration of the $^5$H nucleus. In
particular, the structure of the Pauli-allowed states of the $^5$H nucleus was revealed to correspond to the
subsequent decay $^5$H$\rightarrow n+^4$H$\rightarrow n+n+^3$H.

\appendix

 \section{The Kravchuk polynomials}
 \label{app:1}

The Kravchuk polynomial ${\cal K}^{(p)}_{k}(m)$ of a discrete variable $m$, specified on the interval $0\leq
m\leq\nu$ and orthogonal with weighting function $\rho_{m}$ and norm $d_{k}$
\begin{eqnarray*}
\rho_m={\nu!\over m!(\nu-m)!},~~d_k=\sqrt{{\nu!\over k!(\nu-k)!}}(pq)^{k/2},~~q=1-p,
\end{eqnarray*}
can be defined via hypergeometric function:
\begin{eqnarray*}
\bar{{\cal K}}^{(p)}_{k}(m)\equiv{\cal K}^{(p)}_k(m)\cdot{\sqrt{\rho_m}\over d_k}=\\
= \left\{
\begin{array}{r}
{(-1)^k\over(\nu-m-k)!}\sqrt{{(\nu-k)!(\nu-m)!\over m!k!}}\cdot p^{m+k\over 2}q^{\nu-m-k\over2}
{_2}F_1\left(-k,-m;\nu-m-k+1;-{q\over p}\right), \\ {\rm if} ~~ m\leq\nu-k\vspace{0.5cm}\\
{(-1)^{m-\nu}\over(-\nu+m+k)!}\sqrt{{m!k!\over (\nu-k)!(\nu-m)!}}\cdot q^{m+k-\nu\over 2}p^{\nu-{m+k\over2}}
{_2}F_1\left(-\nu+k,-\nu+m;-\nu+m+k+1;-{q\over p}\right),\\ {\rm if} ~~ m>\nu-k.
\end{array}\right.
\end{eqnarray*}
Note that $\nu$, $m$ and $k$ are the natural integers, $0\leq k\leq \nu.$

Relation of orthogonality for the Kravchuk polynomials looks like:
\begin{eqnarray*}
\sum_{m=0}^{\nu}\bar{{\cal K}}^{(p)}_{k}(m)\bar{{\cal K}}^{(p)}_{\tilde{k}}(m)=\delta_{k,\tilde{k}},~~
\sum_{k=0}^{\nu}\bar{{\cal K}}^{(p)}_{k}(m)\bar{{\cal K}}^{(p)}_{k}(\tilde{m})=\delta_{m,\tilde{m}}.
\end{eqnarray*}
Symmetry relation can be written in the following form:
\begin{eqnarray*}
(-1)^{m+k}{\cal K}^{(p)}_k(m)\cdot{\sqrt{\rho_{m}}\over d_k}= {\cal
K}^{(p)}_{\nu-k}(\nu-m)\cdot{\sqrt{\rho_{\nu-m}}\over d_{\nu-k}}.
\end{eqnarray*}
The Kravchuk polynomial has the limit
\begin{eqnarray}
\label{Hermit} \lim_{\nu\to\infty}{\cal K}^{(p)}_{k}(m){\sqrt{\rho_{m}}\over d_k}={1\over\sqrt{2^k}k!\sqrt{\pi
2pq\nu}}\mbox{H}_k\left({m-p\nu\over\sqrt{2pq\nu}}\right)\exp\left\{-{(m-p\nu)^2\over 4pq\nu}\right\},
\end{eqnarray}
where $\mbox{H}_k(x)$ is the Hermitian polynomial.

\end{document}